\documentclass[
]{ceurart}

\usepackage{listings}
\usepackage{soul}

\begin{document}

\copyrightyear{2025}
\copyrightclause{Copyright for this paper by its authors.
  Use permitted under Creative Commons License Attribution 4.0
  International (CC BY 4.0).}

\conference{CHIRP 2025: Transforming HCI Research in the Philippines Workshop, May 08, 2025, Baguio, Benguet, Philippines}

\title{Designing RoutScape: Geospatial Prototyping with XR for Flood Evacuation Planning}


%
\author[1]{Johndayll Lewis Arizala}[%
email=johndayll_arizala@dlsu.edu.ph
]
\cormark[1]
\fnmark[1]
\address[1]{Human-X Interactions Lab (De La Salle University),
  2401 Taft Ave, Malate, Manila, 1004 Metro Manila}

\author[1]{Joshua Permito}[%
email=research@dwigoric.dev
]
\cormark[1]
\fnmark[1]

\author[1]{Steven Errol Escopete}[%
email=steven_escopete@dlsu.edu.ph
]
\cormark[1]
\fnmark[1]

\author[1]{John Kovie Niño}[%
email=hello@thekovie.com
]
\cormark[1]
\fnmark[1]

\author[1]{Jordan Aiko Deja}[%
email=jordan.deja@dlsu.edu.ph
]
\cormark[1]

\cortext[1]{Corresponding author.}
\fntext[1]{These authors contributed equally.}


\begin{abstract}
Flood response planning in local communities is often hindered by fragmented communication across Disaster Risk Reduction and Management (DRRM) councils. In this work, we explore how extended reality (XR) can support more effective planning through narrative-driven design. We present Routscape, an XR prototype for visualizing flood scenarios and evacuation routes, developed through iterative prototyping and user-centered design with DRRM officers. By grounding the system in real-world experiences and localized narratives, we highlight how XR can aid in fostering shared understanding and spatial sensemaking in disaster preparedness efforts.
\end{abstract}

\begin{keywords}
  extended reality \sep
  iterative prototyping \sep
  flood evacuation \sep
  flood planning \sep
  flood preparedness \sep
  disaster risk reduction
\end{keywords}

\maketitle
\section{Introduction and Background}
\par The Philippines faces heightened exposure to natural hazards, especially typhoons, making disaster risk reduction a national priority. The National Disaster Risk Reduction Management Plan emphasizes the importance of risk assessments as a foundational step toward preparedness. Key stakeholders in flood evacuation planning rely on structured frameworks for risk assessment \cite{ravago2020localized, mercado2021fuzzy, robielos2020development, walsh2020measuring}, alongside evaluations of flood risk and evacuation centers \cite{paranaque_contingency_2015, gacu2022flood, cajucom2019evaluation}. These tools form the basis for proactive planning strategies.

\par However, the implementation of disaster preparedness remains hindered by coordination challenges across institutions \cite{ndrrmp_2020}. For instance, during Typhoon Ketsana (local: Tropical Storm Ondoy), the Philippine National Capital Region was both affected and responsible for providing aid, illustrating the blurred boundaries between victims and responders. In the case of Severe Tropical Storm Washi (local: Severe Tropical Storm Sendong), recommended pre-emptive evacuation measures were not carried out.

\par We have seen multiple works aimed at simulating flood data \cite{chua_vizrisk_2019, up_resilience_institute_noah_2021, yan_review_2015, nunes_augmented_2018, wang_development_2016}, using Virtual Reality (VR) in disaster response training \cite{sharma_immersive_2014, hsu_state_2013, nguyen_vrescuer_2019}, and demonstrating the capabilities of Mixed Reality (MR) devices while implementing collaborative features \cite{rydvanskiy_mixed_2021}. There is also a similar work using XR for collaborative wildfire evacuation planning \cite{ghosh_designing_2024}. Much of our work was derived from the work of \citeauthor{rydvanskiy_mixed_2021}~\cite{rydvanskiy_mixed_2021}, which focused on collaborative MR for flood simulation.

\par Effective flood preparedness requires coordinated planning across multiple actors \cite{ndrrmp_2020}. \citeauthor{ghosh_designing_2024}~\cite{ghosh_designing_2024} mentioned that XR supports collaborative ``immersive sensemaking.'' We define \textit{visualization} and \textit{collaboration} as key features for \textit{collaborative sensemaking}. The works we have seen aid in both of these features. Thus, we see an opportunity for XR to support coordinated planning by offering accessible, real-time tools for collaborative response—a key indicator of preparedness \cite{ravago2020localized, mercado2021fuzzy, prasetyo2020confirmatory}. XR has been shown to facilitate collaboration in complex spatial tasks \cite{sereno_collaborative_2022}, yet its application in flood evacuation planning involving multiple stakeholders remains largely unexplored.

\par In this work, we present RoutScape, a synchronous collaborative XR prototype developed through iterative prototyping and user-centered design. Drawing from real-world flood scenarios and grounded in user feedback, we explore how narrative-driven XR design can support communication and planning in disaster contexts. Our contributions are the initial insights drawn from the user testing we have conducted to date.
\section{Iterative Prototyping of RoutScape}

\par We conducted two main phases: (1)~development and iterative prototyping of a proof-of-concept~(PoC), and (2)~refinement and iterative prototyping of \textit{RoutScape}.

\par
\begin{figure}[]
    \centering
    \includegraphics[width=0.49\linewidth]{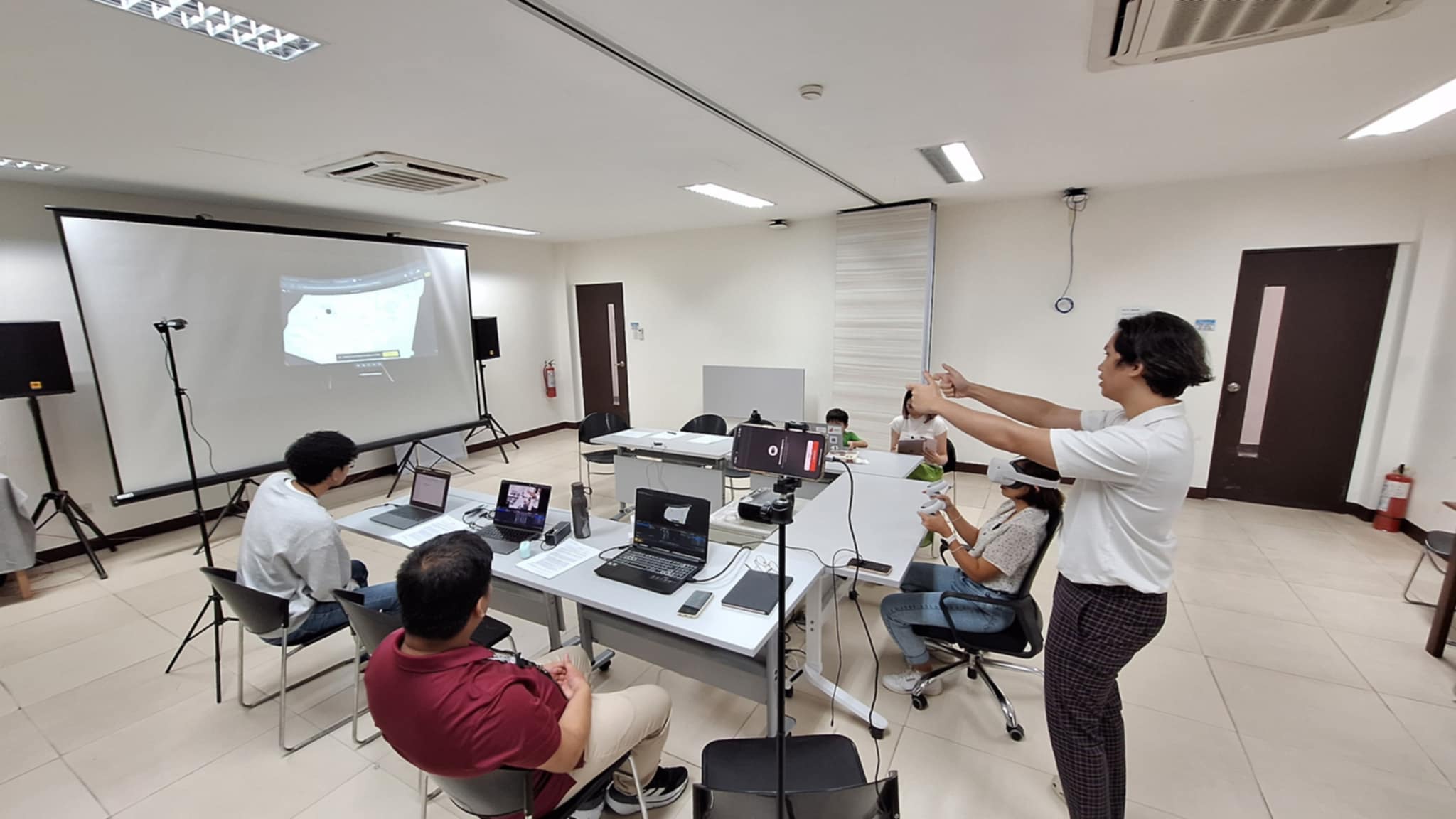}
    \includegraphics[width=0.49\linewidth]{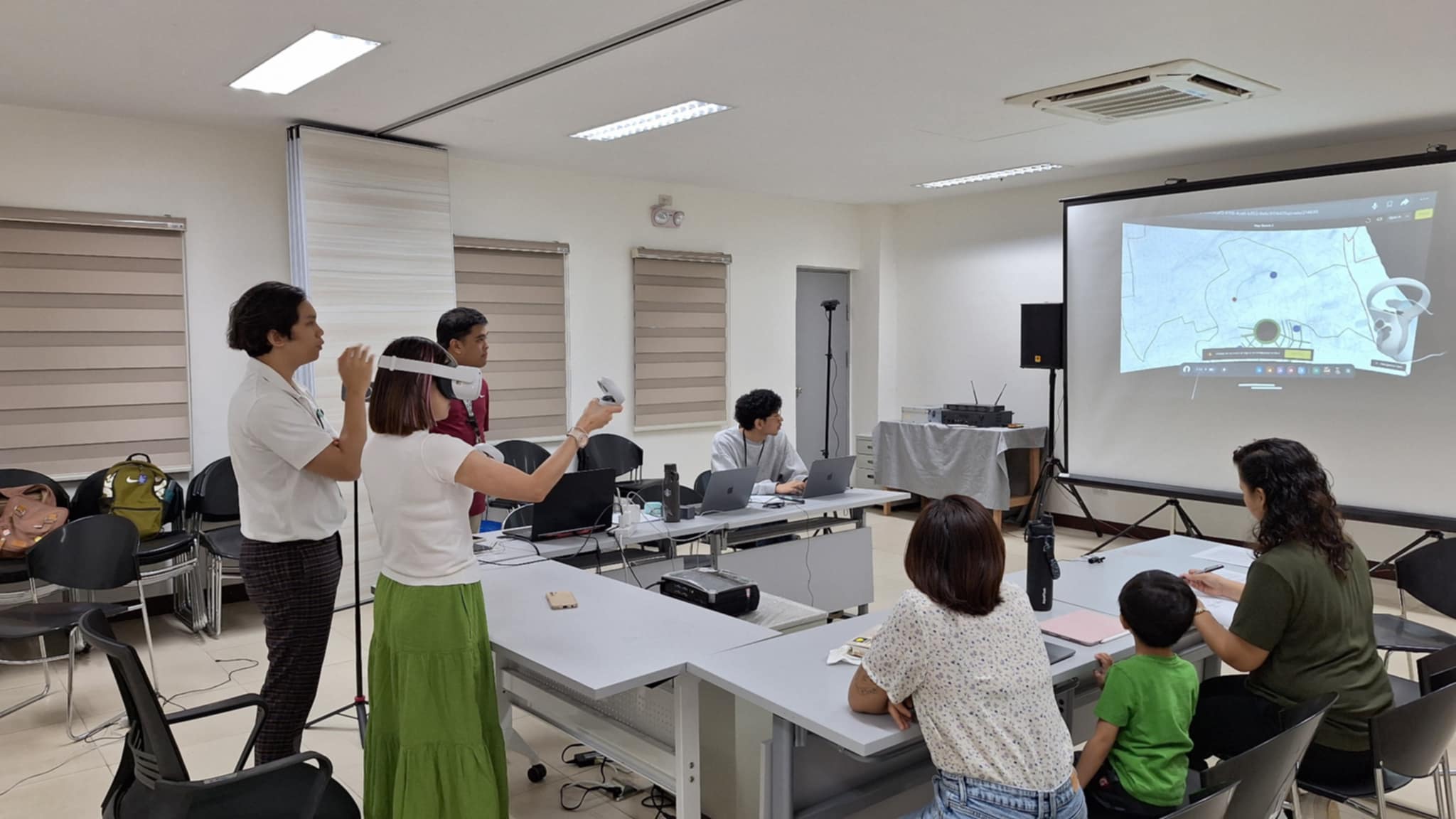}
    \caption{Overview of the XR testing setup showing camera placements and user play area.}
    \label{fig:setup_poc_testing}
\end{figure}

 All iterations within the two phases followed a similar workflow: prioritizing specific features, building interactive prototypes, conducting expert-led testing sessions, and refining the design based on the feedback collected. Moreover, we created a setup that applies to both phases where there is a) a camera that captures the movements of the participants, b) a lapel microphone attached to all participants, and c) a projector displaying the point of view of the headset (see Figure \ref{fig:setup_poc_testing}). We partnered with experts in geospatial data visualization as well as disaster risk reduction management officers, in which none of the participants had prior hands on experience in XR technologies. Lastly, demographic data, such as the participant's age, was not formally recorded. They have actively participated in each iteration, offering valuable feedback and expertise in refining the features of \textit{RoutScape}. 

\subsection{RoutScape Proof-of-Concept}

\par
\begin{figure}[]
  \centering
  \includegraphics[width=0.49\linewidth]{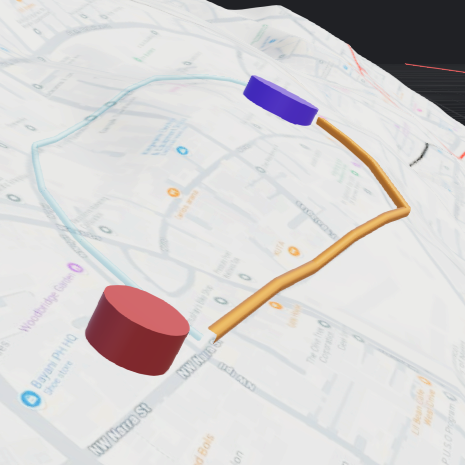}
  \includegraphics[width=0.49\linewidth]{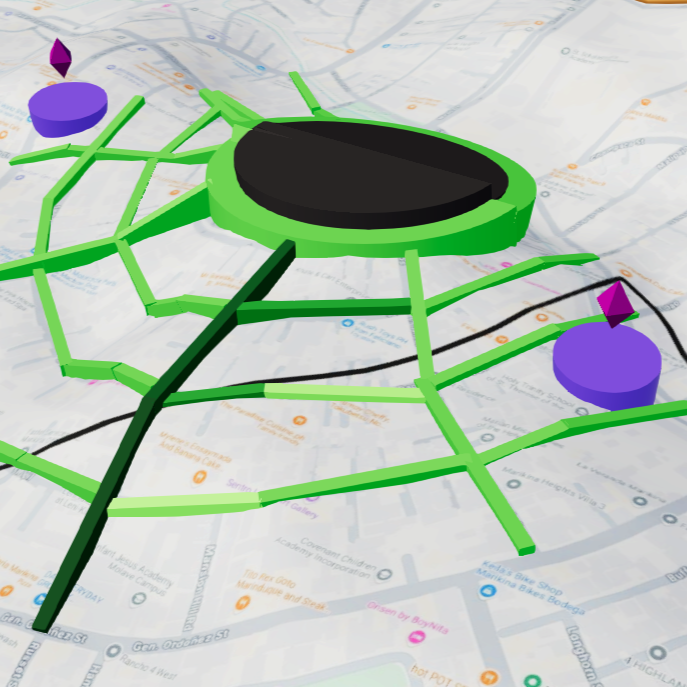}\\
  \includegraphics[width=0.49\linewidth]{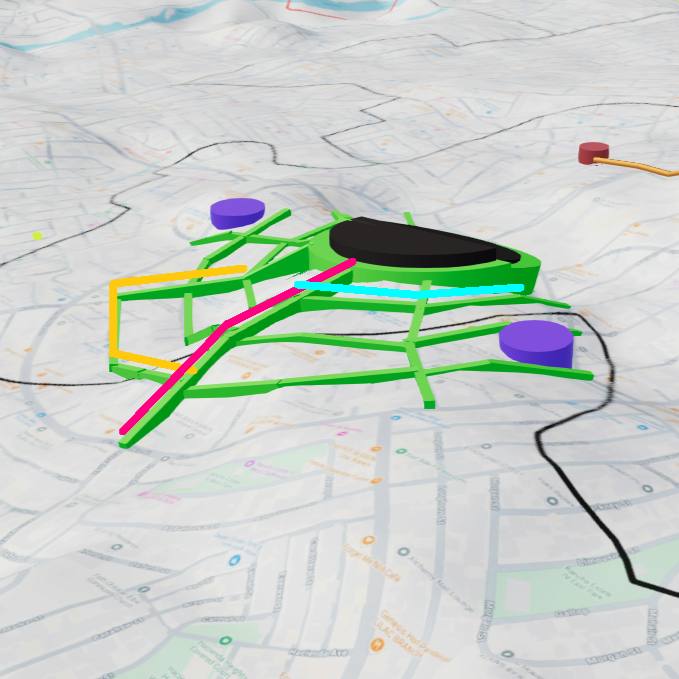}
  \includegraphics[width=0.49\linewidth]{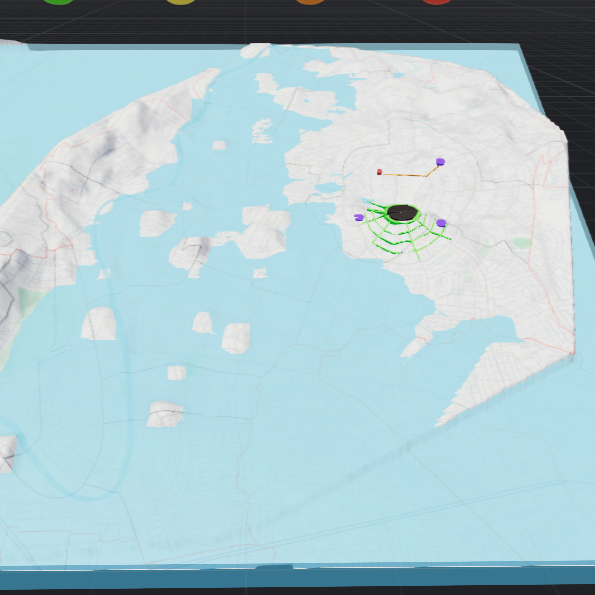}
  \caption{Features: Navigating the Map (top-left), Plotting Pins (top-right), Drawing Routes (bottom-left), Flood Visualization (bottom-right).}
  \label{fig:bezi}
\end{figure}

 We implemented a proof-of-concept~(PoC) prototype using \textit{Bezi}, with its initial features inspired by previous works \cite{chua_vizrisk_2019, rydvanskiy_mixed_2021, ghosh_designing_2024}. This initial version was designed for single-user interaction, focusing on key visualizations and interaction techniques for flood evacuation planning. Here, the user is shown a god's eye view of the map; they are able to pan, zoom, and rotate, visualize different flood levels, plot a pin on specific locations, and draw routes (see Figure \ref{fig:bezi}). Each session followed a set of steps where 1) the identified priority features are demonstrated, 2) the prototype is used by the participants one at a time, and 3) feedback is gathered through questions formed by the authors throughout the session. Participants used a \textit{Meta Quest 2} headset to interact with the PoC, using the controller instead of hand gestures. Insights from these sessions informed the development of the next version of \textit{RoutScape}.

\subsection{RoutScape Prototype}

\par
\begin{figure}[]
  \centering
  \includegraphics[width=0.49\linewidth]{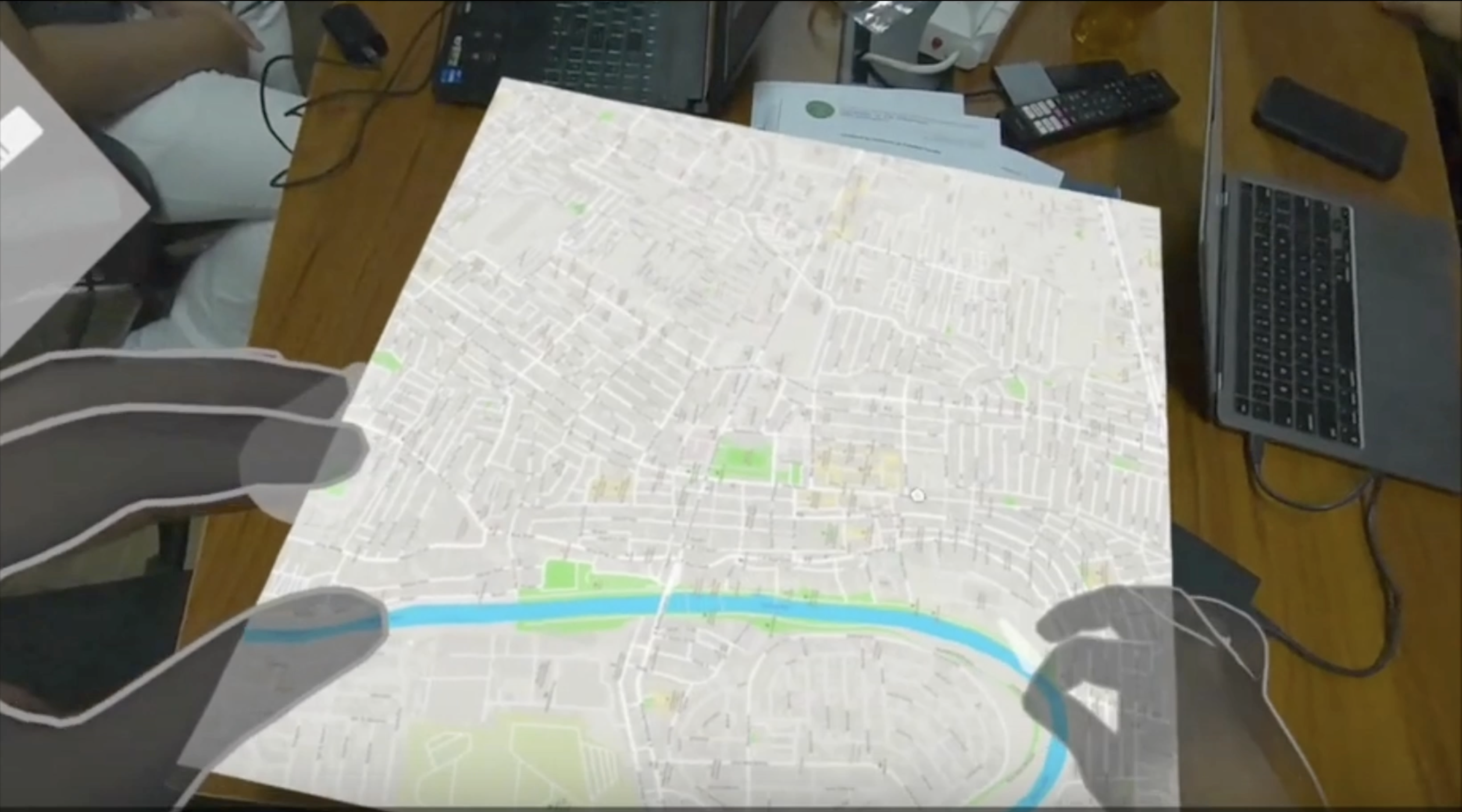}
  \includegraphics[width=0.49\linewidth]{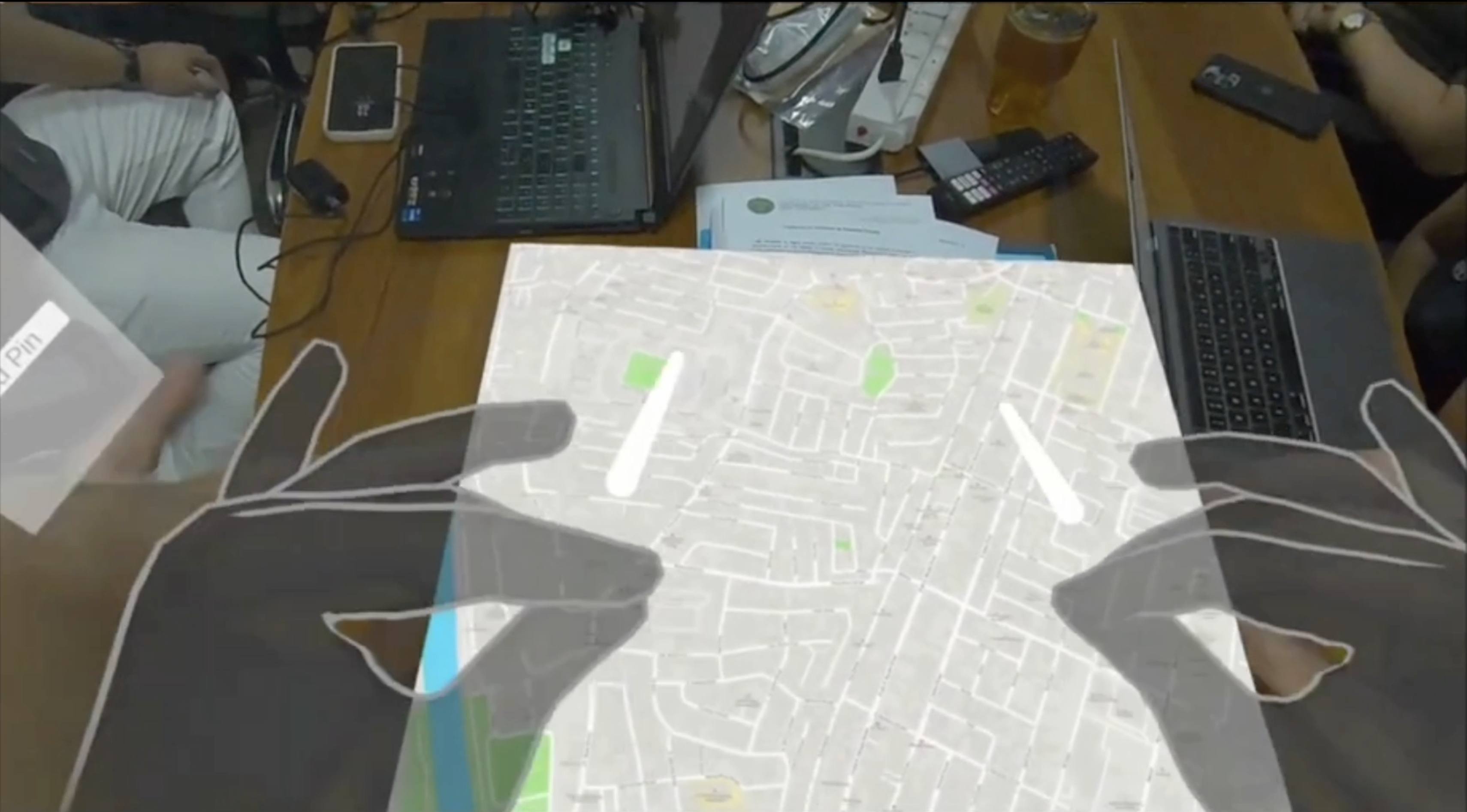}\\
  \includegraphics[width=0.49\linewidth]{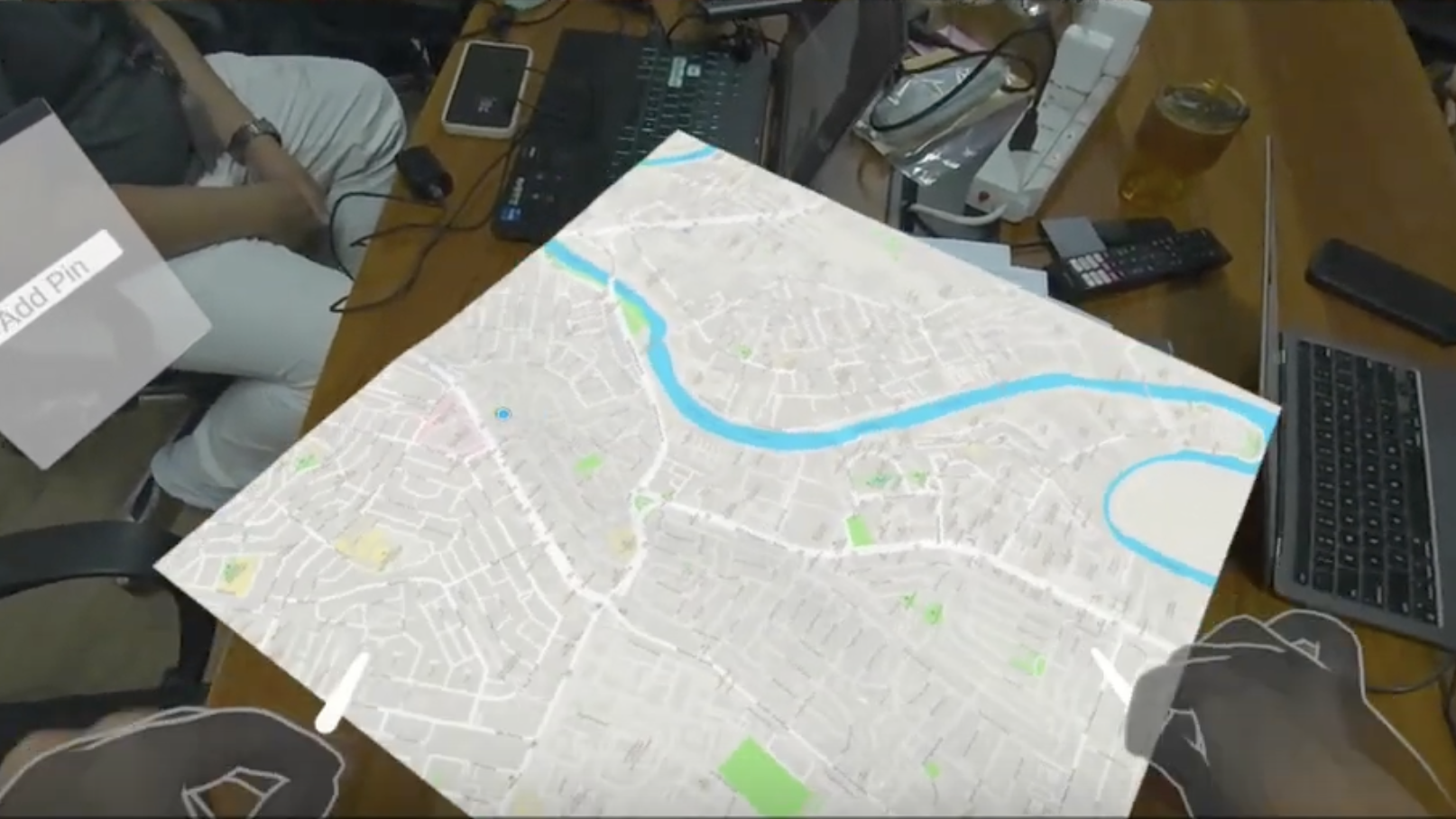}
  \includegraphics[width=0.49\linewidth]{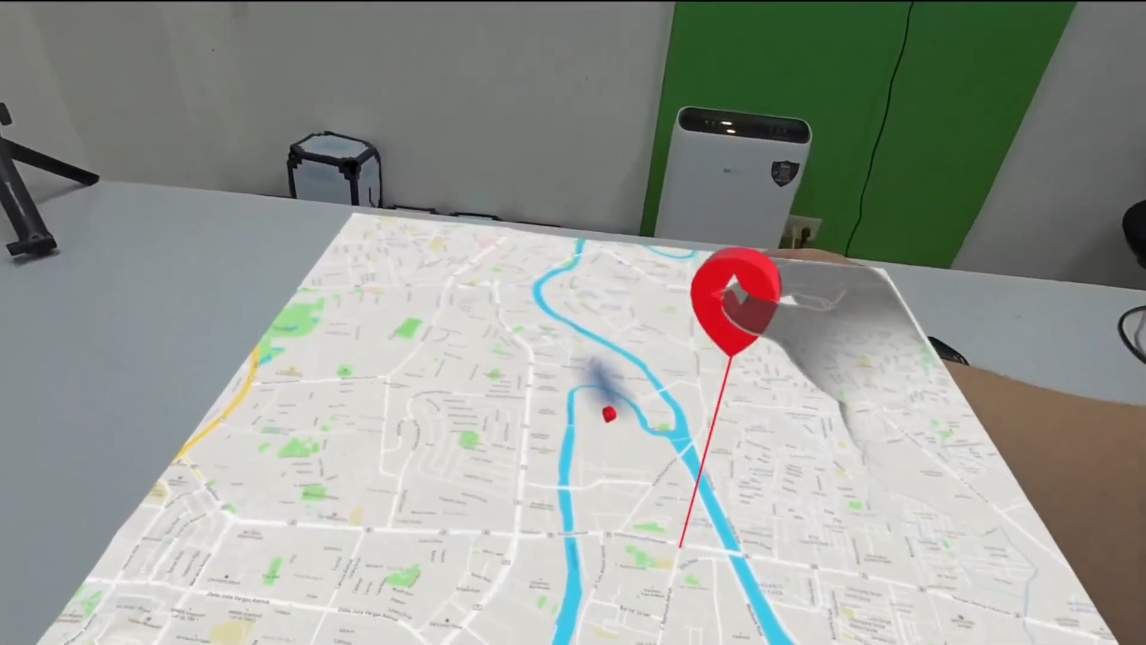}
  \includegraphics[width=0.49\linewidth]{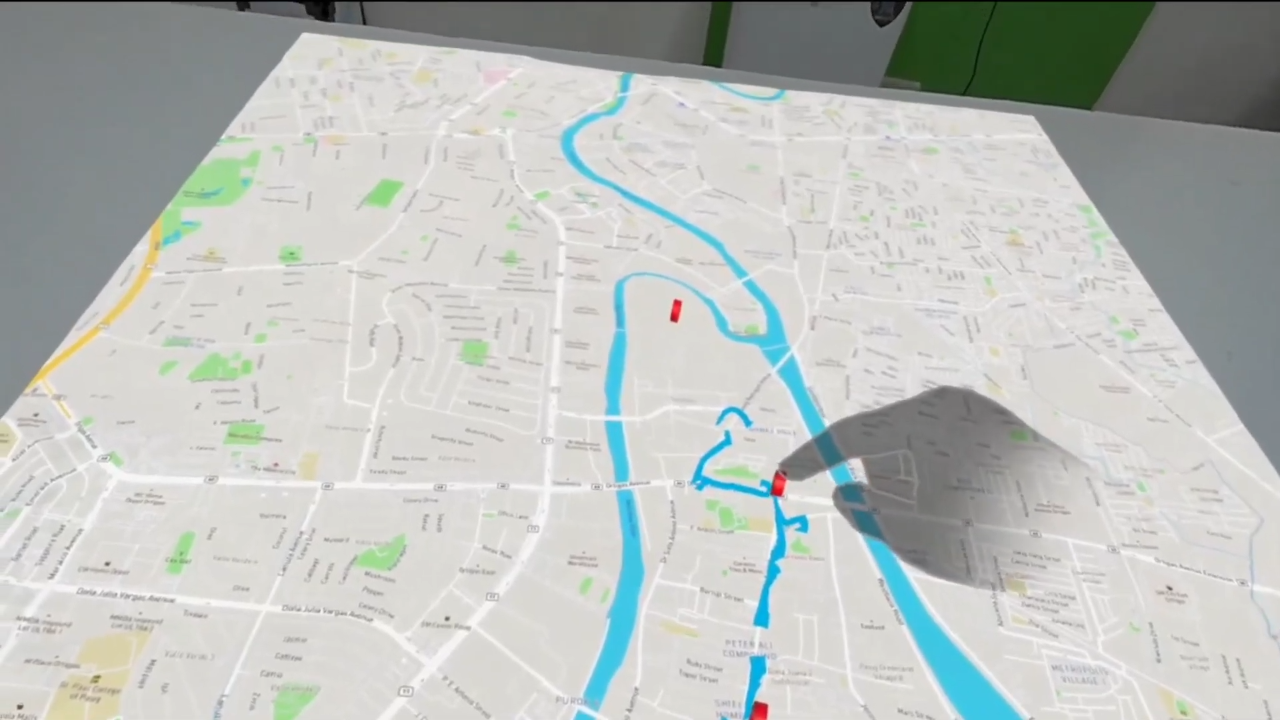}
  \includegraphics[width=0.49\linewidth]{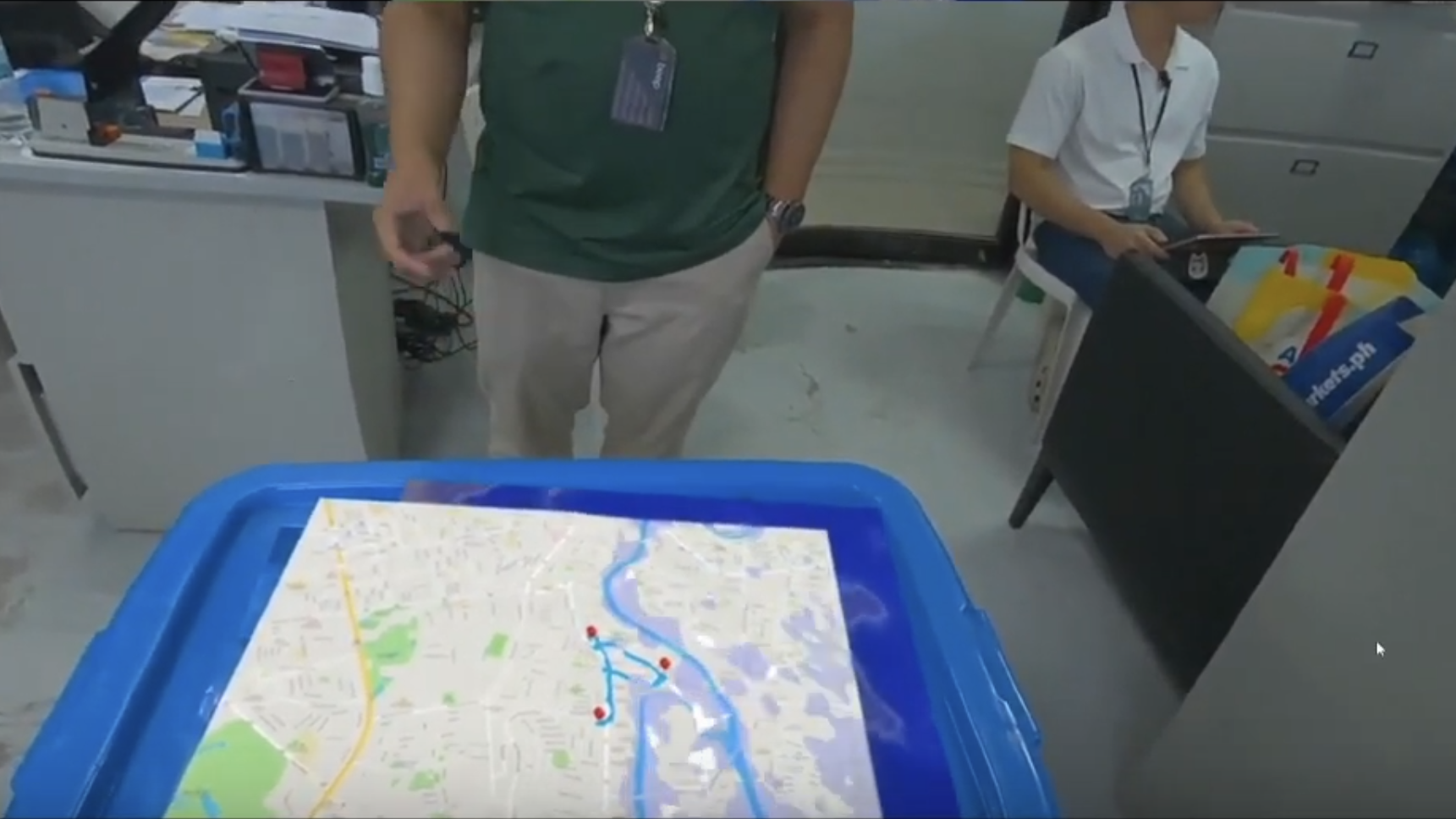}
  \caption{Features: Panning (top-left), Zooming (top-right), Rotating (middle-left), Plotting (middle-right), Routing (bottom-left), Flood Visualization (bottom-right).} 
  \label{fig:interactions}
\end{figure}

 Building upon our findings and insights from the previous phase, we refined the core features to develop \textit{RoutScape}. This is developed in the \textit{Unity} game engine, using the \textit{Meta XR All-in-One SDK}. While the core features of the PoC was preserved (see Figure \ref{fig:interactions}), this version employed hand gestures instead of controllers and, moreover, supported multi-user interaction. Each session followed a similar flow from the previous phase, but with the following changes: 1) strictly two participants are synchronously interacting with the prototype, and 2) instead of freely interacting the prototype, a simple set of tasks are given for the participants to accomplish. We used a \textit{Meta Quest 2} and \textit{Meta Quest 3} headset to accommodate two participants. 

\section{Insights Learned and Discussion}

\par Our evaluation of the PoC with DRRM officers and geospatial experts underscored the critical role of visualization in flood evacuation planning. DRRM officers emphasized the need to integrate contextual data—such as flood hazard zones—directly into the visual interface to support more informed decision-making. Meanwhile, geospatial experts focused primarily on gesture design and interaction techniques. Since the PoC supported only single-user interaction, collaboration-related feedback was not captured during this phase. These findings informed the design and development of our multi-user prototype, \textit{RoutScape}.

\par Subsequent iterations with \textit{RoutScape} reaffirmed the importance of effective data visualization. Participants highlighted the need for clearly legible text on the map, as well as displaying point-of-interests similar to Google Maps; which helped users orient themselves within the spatial context. Visualizing critical point facilities—such as evacuation centers; as well as a smoother route-drawing experience—was seen as essential for route planning. Furthermore, both DRRM officers and geospatial experts stressed the importance of representing flood depth to properly assess risk across different areas. The DRRM officers suggested visualizing the flood depth similar to a flood hazard map, where different colors represent either the hazard level or depth.

\par A notable limitation emerged that was emphasized by all experts: the cost and accessibility of head-mounted displays (HMDs), which may constrain broader deployment of XR-based planning tools in local government settings.

\section{Conclusion and Future Work}
\par In this paper, we introduced \textit{RoutScape}, an XR tool designed to support synchronous, collaborative flood evacuation planning. Through iterative prototyping with DRRM officers and geospatial experts, we identified essential features for effective geospatial visualization, gesture-based interaction, and spatial orientation in immersive environments. Our user-centered approach revealed how narrative and domain-specific feedback can shape XR tools for disaster preparedness.

\par Moving forward, we aim to further develop \textit{RoutScape} by incorporating insights from our design iterations. Future work includes:
a)~conducting detailed evaluations of individual features with domain experts to assess their utility and usability, and
b)~engaging stakeholders in structured feedback sessions using a more complete prototype to explore additional use cases and refine the system for practical deployment.

\begin{acknowledgments} We thank the University of the Philippines Resilience Institute (UPRI) Project NOAH\footnotemark[1] team, as well as the Bacoor and Pasig City Disaster Risk Reduction and Management Offices (CDRRMO), for their valuable feedback and insights that guided the development of \textit{RoutScape}.
\end{acknowledgments}

\footnotetext[1]{Project NOAH Website: \url{https://noah.up.edu.ph}}
\section*{Declaration on Generative AI}
 Generative AI tools were used in the preparation of this paper to improve grammar, clarity, and language flow. All core ideas, research contributions, analysis, and writing were authored by the researchers.
  \newline
  
\bibliography{ceur}

\begin{thebibliography}{20}
\expandafter\ifx\csname natexlab\endcsname\relax\def\natexlab#1{#1}\fi
\providecommand{\url}[1]{\texttt{#1}}
\providecommand{\href}[2]{#2}
\providecommand{\path}[1]{#1}
\providecommand{\DOIprefix}{doi:}
\providecommand{\ArXivprefix}{arXiv:}
\providecommand{\URLprefix}{URL: }
\providecommand{\Pubmedprefix}{pmid:}
\providecommand{\doi}[1]{\href{http://dx.doi.org/#1}{\path{#1}}}
\providecommand{\Pubmed}[1]{\href{pmid:#1}{\path{#1}}}
\providecommand{\bibinfo}[2]{#2}
\ifx\xfnm\relax \def\xfnm[#1]{\unskip,\space#1}\fi
\bibitem[{Ravago et~al.(2020)Ravago, Mapa, Aycardo, and Abrigo}]{ravago2020localized}
\bibinfo{author}{M.-L.~V. Ravago}, \bibinfo{author}{C.~D.~S. Mapa}, \bibinfo{author}{A.~G. Aycardo}, \bibinfo{author}{M.~R. Abrigo},
\newblock \bibinfo{title}{Localized disaster risk management index for the philippines: Is your municipality ready for the next disaster?},
\newblock \bibinfo{journal}{International Journal of Disaster Risk Reduction} \bibinfo{volume}{51} (\bibinfo{year}{2020}) \bibinfo{pages}{101913}.
\bibitem[{Mercado et~al.(2021)Mercado, Kawamura, Amaguchi, and Rubio}]{mercado2021fuzzy}
\bibinfo{author}{J.~M.~R. Mercado}, \bibinfo{author}{A.~Kawamura}, \bibinfo{author}{H.~Amaguchi}, \bibinfo{author}{C.~J.~P. Rubio},
\newblock \bibinfo{title}{Fuzzy based multi-criteria m\&e of the integrated flood risk management performance using priority ranking methodology: A case study in metro manila, philippines},
\newblock \bibinfo{journal}{International Journal of Disaster Risk Reduction} \bibinfo{volume}{64} (\bibinfo{year}{2021}) \bibinfo{pages}{102498}.
\bibitem[{Robielos et~al.(2020)Robielos, Lin, Senoro, and Ney}]{robielos2020development}
\bibinfo{author}{R.~A.~C. Robielos}, \bibinfo{author}{C.~J. Lin}, \bibinfo{author}{D.~B. Senoro}, \bibinfo{author}{F.~P. Ney},
\newblock \bibinfo{title}{Development of vulnerability assessment framework for disaster risk reduction at three levels of geopolitical units in the philippines},
\newblock \bibinfo{journal}{Sustainability} \bibinfo{volume}{12} (\bibinfo{year}{2020}) \bibinfo{pages}{8815}.
\bibitem[{Walsh and Hallegatte(2020)}]{walsh2020measuring}
\bibinfo{author}{B.~Walsh}, \bibinfo{author}{S.~Hallegatte},
\newblock \bibinfo{title}{Measuring natural risks in the philippines: Socioeconomic resilience and wellbeing losses},
\newblock \bibinfo{journal}{Economics of Disasters and Climate Change} \bibinfo{volume}{4} (\bibinfo{year}{2020}) \bibinfo{pages}{249--293}.
\bibitem[{{Paranaque City DRRM Office}(2015)}]{paranaque_contingency_2015}
\bibinfo{author}{{Paranaque City DRRM Office}}, \bibinfo{title}{{Paranaque} {Contingency Plan}}, \bibinfo{year}{2015}. \URLprefix \url{https://carbonn.org/uploads/tx_carbonndata/CP%20FLOOD%20FINAL%20DRAFT%202015%20(2)%20(1).pdf#page=91.09}.
\bibitem[{Gacu et~al.(2022)Gacu, Monjardin, Senoro, and Tan}]{gacu2022flood}
\bibinfo{author}{J.~G. Gacu}, \bibinfo{author}{C.~E.~F. Monjardin}, \bibinfo{author}{D.~B. Senoro}, \bibinfo{author}{F.~J. Tan},
\newblock \bibinfo{title}{Flood risk assessment using gis-based analytical hierarchy process in the municipality of odiongan, romblon, philippines},
\newblock \bibinfo{journal}{Applied Sciences} \bibinfo{volume}{12} (\bibinfo{year}{2022}) \bibinfo{pages}{9456}.
\bibitem[{Cajucom et~al.(2019)Cajucom, Chao~Jr, Constantino, Ejares, Quillope, Solomon, and Ringor}]{cajucom2019evaluation}
\bibinfo{author}{E.~Cajucom}, \bibinfo{author}{G.~Chao~Jr}, \bibinfo{author}{G.~Constantino}, \bibinfo{author}{J.~Ejares}, \bibinfo{author}{S.~Quillope}, \bibinfo{author}{H.~Solomon}, \bibinfo{author}{C.~Ringor},
\newblock \bibinfo{title}{Evaluation of the spatial distribution of evacuation centers in metro manila, philippines},
\newblock \bibinfo{journal}{The International Archives of the Photogrammetry, Remote Sensing and Spatial Information Sciences} \bibinfo{volume}{42} (\bibinfo{year}{2019}) \bibinfo{pages}{79--85}.
\bibitem[{{Office of Civil Defense}(2020)}]{ndrrmp_2020}
\bibinfo{author}{{Office of Civil Defense}}, \bibinfo{title}{NATIONAL DISASTER RISK REDUCTION AND MANAGEMENT PLAN (NDRRMP) 2020 – 2030}, \bibinfo{edition}{v5} ed., \bibinfo{publisher}{Office of Civil Defense}, \bibinfo{address}{Quezon City, Philippines}, \bibinfo{year}{2020}.
\bibitem[{Chua and Samson(2019)}]{chua_vizrisk_2019}
\bibinfo{author}{U.~Chua}, \bibinfo{author}{B.~P. Samson}, \bibinfo{title}{\#{VizRisk} — {Flooding} in {Marikina} {City}: {A} {Case} {Study}}, \bibinfo{year}{2019}. \URLprefix \url{https://medium.com/dlsu-comet/vizrisk-flooding-in-marikina-city-a-case-study-2a59cf0dd1ba}.
\bibitem[{{UP Resilience Institute}(2021)}]{up_resilience_institute_noah_2021}
\bibinfo{author}{{UP Resilience Institute}}, \bibinfo{title}{{NOAH} {Center}}, \bibinfo{year}{2021}. \URLprefix \url{https://noahcenter.up.edu.ph/}.
\bibitem[{Yan et~al.(2015)Yan, Di~Baldassarre, Solomatine, and Schumann}]{yan_review_2015}
\bibinfo{author}{K.~Yan}, \bibinfo{author}{G.~Di~Baldassarre}, \bibinfo{author}{D.~P. Solomatine}, \bibinfo{author}{G.~J.-P. Schumann},
\newblock \bibinfo{title}{A review of low-cost space-borne data for flood modelling: topography, flood extent and water level},
\newblock \bibinfo{journal}{Hydrological Processes} \bibinfo{volume}{29} (\bibinfo{year}{2015}) \bibinfo{pages}{3368--3387}. \URLprefix \url{https://onlinelibrary.wiley.com/doi/abs/10.1002/hyp.10449}. \DOIprefix\doi{10.1002/hyp.10449}, \bibinfo{note}{\_eprint: https://onlinelibrary.wiley.com/doi/pdf/10.1002/hyp.10449}.
\bibitem[{Nunes et~al.(2018)Nunes, Lucas, Simões-Marques, and Correia}]{nunes_augmented_2018}
\bibinfo{author}{I.~L. Nunes}, \bibinfo{author}{R.~Lucas}, \bibinfo{author}{M.~Simões-Marques}, \bibinfo{author}{N.~Correia},
\newblock \bibinfo{title}{Augmented {Reality} in {Support} of {Disaster} {Response}},
\newblock in: \bibinfo{editor}{I.~L. Nunes} (Ed.), \bibinfo{booktitle}{Advances in {Human} {Factors} and {Systems} {Interaction}}, \bibinfo{publisher}{Springer International Publishing}, \bibinfo{address}{Cham}, \bibinfo{year}{2018}, pp. \bibinfo{pages}{155--167}. \DOIprefix\doi{10.1007/978-3-319-60366-7_15}.
\bibitem[{Wang et~al.(2016)Wang, Jiang, Xie, Miller, and Brown}]{wang_development_2016}
\bibinfo{author}{C.~Wang}, \bibinfo{author}{Y.~Jiang}, \bibinfo{author}{H.~Xie}, \bibinfo{author}{D.~Miller}, \bibinfo{author}{I.~Brown},
\newblock \bibinfo{title}{Development of a {Flood} {Warning} {Simulation} {System}: {A} {Case} {Study} of 2007 {Tewkesbury} {Flood}},
\newblock \bibinfo{journal}{E3S Web Conf.} \bibinfo{volume}{7} (\bibinfo{year}{2016}) \bibinfo{pages}{18021}. \URLprefix \url{http://www.e3s-conferences.org/10.1051/e3sconf/20160718021}. \DOIprefix\doi{10.1051/e3sconf/20160718021}.
\bibitem[{Sharma et~al.(2014)Sharma, Jerripothula, Mackey, and Soumare}]{sharma_immersive_2014}
\bibinfo{author}{S.~Sharma}, \bibinfo{author}{S.~Jerripothula}, \bibinfo{author}{S.~Mackey}, \bibinfo{author}{O.~Soumare},
\newblock \bibinfo{title}{Immersive virtual reality environment of a subway evacuation on a cloud for disaster preparedness and response training},
\newblock in: \bibinfo{booktitle}{2014 {IEEE} {Symposium} on {Computational} {Intelligence} for {Human}-like {Intelligence} ({CIHLI})}, \bibinfo{year}{2014}, pp. \bibinfo{pages}{1--6}. \URLprefix \url{https://ieeexplore.ieee.org/abstract/document/7013380}. \DOIprefix\doi{10.1109/CIHLI.2014.7013380}.
\bibitem[{Hsu et~al.(2013)Hsu, Li, Bayram, Levinson, Yang, and Monahan}]{hsu_state_2013}
\bibinfo{author}{E.~B. Hsu}, \bibinfo{author}{Y.~Li}, \bibinfo{author}{J.~D. Bayram}, \bibinfo{author}{D.~Levinson}, \bibinfo{author}{S.~Yang}, \bibinfo{author}{C.~Monahan},
\newblock \bibinfo{title}{State of {Virtual} {Reality} {Based} {Disaster} {Preparedness} and {Response} {Training}},
\newblock \bibinfo{journal}{PLoS Curr} \bibinfo{volume}{5} (\bibinfo{year}{2013}) \bibinfo{pages}{ecurrents.dis.1ea2b2e71237d5337fa53982a38b2aff}. \URLprefix \url{https://www.ncbi.nlm.nih.gov/pmc/articles/PMC3644293/}. \DOIprefix\doi{10.1371/currents.dis.1ea2b2e71237d5337fa53982a38b2aff}.
\bibitem[{Nguyen et~al.(2019)Nguyen, Jung, and Dang}]{nguyen_vrescuer_2019}
\bibinfo{author}{V.~T. Nguyen}, \bibinfo{author}{K.~Jung}, \bibinfo{author}{T.~Dang},
\newblock \bibinfo{title}{{VRescuer}: {A} {Virtual} {Reality} {Application} for {Disaster} {Response} {Training}},
\newblock in: \bibinfo{booktitle}{2019 {IEEE} {International} {Conference} on {Artificial} {Intelligence} and {Virtual} {Reality} ({AIVR})}, \bibinfo{year}{2019}, pp. \bibinfo{pages}{199--1993}. \URLprefix \url{https://ieeexplore.ieee.org/abstract/document/8942307}. \DOIprefix\doi{10.1109/AIVR46125.2019.00042}.
\bibitem[{Rydvanskiy and Hedley(2021)}]{rydvanskiy_mixed_2021}
\bibinfo{author}{R.~Rydvanskiy}, \bibinfo{author}{N.~Hedley},
\newblock \bibinfo{title}{Mixed {Reality} {Flood} {Visualizations}: {Reflections} on {Development} and {Usability} of {Current} {Systems}},
\newblock \bibinfo{journal}{IJGI} \bibinfo{volume}{10} (\bibinfo{year}{2021}) \bibinfo{pages}{82}. \URLprefix \url{https://www.mdpi.com/2220-9964/10/2/82}. \DOIprefix\doi{10.3390/ijgi10020082}.
\bibitem[{Ghosh et~al.(2024)Ghosh, Wang, Zhou, Lin, Mcveigh-Schultz, and Isbister}]{ghosh_designing_2024}
\bibinfo{author}{S.~Ghosh}, \bibinfo{author}{Y.~Wang}, \bibinfo{author}{W.~Zhou}, \bibinfo{author}{K.~Lin}, \bibinfo{author}{J.~Mcveigh-Schultz}, \bibinfo{author}{K.~Isbister},
\newblock \bibinfo{title}{Designing {Shared} {VR} {Tools} for {Spatial} {Scientific} {Sensemaking} {About} {Wildfire} {Evacuation}},
\newblock in: \bibinfo{booktitle}{Extended {Abstracts} of the {CHI} {Conference} on {Human} {Factors} in {Computing} {Systems}}, {CHI} {EA} '24, \bibinfo{publisher}{Association for Computing Machinery}, \bibinfo{address}{New York, NY, USA}, \bibinfo{year}{2024}, pp. \bibinfo{pages}{1--5}. \URLprefix \url{https://doi.org/10.1145/3613905.3650819}. \DOIprefix\doi{10.1145/3613905.3650819}.
\bibitem[{Prasetyo et~al.(2020)Prasetyo, Senoro, German, Robielos, and Ney}]{prasetyo2020confirmatory}
\bibinfo{author}{Y.~T. Prasetyo}, \bibinfo{author}{D.~B. Senoro}, \bibinfo{author}{J.~D. German}, \bibinfo{author}{R.~A.~C. Robielos}, \bibinfo{author}{F.~P. Ney},
\newblock \bibinfo{title}{Confirmatory factor analysis of vulnerability to natural hazards: A household vulnerability assessment in marinduque island, philippines},
\newblock \bibinfo{journal}{International Journal of Disaster Risk Reduction} \bibinfo{volume}{50} (\bibinfo{year}{2020}) \bibinfo{pages}{101831}.
\bibitem[{Sereno et~al.(2022)Sereno, Wang, Besançon, McGuffin, and Isenberg}]{sereno_collaborative_2022}
\bibinfo{author}{M.~Sereno}, \bibinfo{author}{X.~Wang}, \bibinfo{author}{L.~Besançon}, \bibinfo{author}{M.~J. McGuffin}, \bibinfo{author}{T.~Isenberg},
\newblock \bibinfo{title}{Collaborative {Work} in {Augmented} {Reality}: {A} {Survey}},
\newblock \bibinfo{journal}{IEEE Transactions on Visualization and Computer Graphics} \bibinfo{volume}{28} (\bibinfo{year}{2022}) \bibinfo{pages}{2530--2549}. \URLprefix \url{https://ieeexplore.ieee.org/abstract/document/9234650}. \DOIprefix\doi{10.1109/TVCG.2020.3032761}, \bibinfo{note}{conference Name: IEEE Transactions on Visualization and Computer Graphics}.

\end{thebibliography}
\appendix
\end{document}